\documentclass[preprint,11pt]{aastex}
\usepackage{amsmath}
\tighten
\newcommand{\asec}{\hbox to 1pt{}\rlap{$^{\prime\prime}$}.\hbox to 2pt{}}
\newcommand{\amin}{\hbox to 1pt{}\rlap{$^{\prime}$}.\hbox to 2pt{}}

\def\HDST{{\it HDST}}
\def\JWST{{\it JWST}}
\def\HST{{\it HST}}
\def\CBLE{{\it Cosmic Births}}

\begin{document}

\title{A Response to Elvis' 2015 Critique of the AURA Report ``From Cosmic Birth to Living Earths" }

\author{Sara Seager}
\affil{Massachusetts Institute of Technology}
\author{Julianne J. Dalcanton}
\affil{University of Washington, Seattle}
\author{Marc Postman, Jason Tumlinson}
\affil{Space Telescope Science Institute}
\author{John C. Mather}
\affil{NASA Goddard Space Flight Center}
\author{On behalf of the entire AURA \HDST\ Report Committee}

\begin{abstract}
To ensure progress in astronomy over the coming decades, the key questions are ``what facilities will we build, and when?'' Toward this end, the Association of Universities for Research in Astronomy (AURA) recently commissioned a study on future space-based options for UV and optical astronomy. The resulting study -- ``From Cosmic Birth to Living Earths" -- concluded that a space telescope equipped with a 12-meter class primary mirror would make fundamental advances across virtually all of astrophysics, including finding and characterizing the atmospheres of dozens of Earth-like planets. This ambitious telescope concept is referred to as the High Definition Space Telescope (\HDST\footnote{See HDSTvision.org}). In a recent arXiv white paper, Elvis (2015) critiqued a subset of the findings of the AURA study, focusing on the detection and characterization of rocky exoplanets in the habitable zone. In this response, we clarify these issues to confirm that \HDST\ would play a transformative role in the study of terrestrial worlds.  Its capabilities for studying exoplanets would be truly unique, even in 2035, and would complement \HDST's broad and deep range of exciting astrophysics. 
\end{abstract}

\section{Introduction: Setting the Record Straight on the \HDST\ Science Case}

The U.S. astronomical community has an exemplary history of carefully evaluating and planning its investments in future shared facilities.  Exercises like the Decadal Survey process have consistently led to projects that help realize the most compelling scientific goals of each generation.  Key to this process is community input, both in developing and evaluating new missions.

Recently, our committee presented one possible mission concept for a
next generation ultraviolet (UV), optical, and near-infrared (NIR)
space telescope, which (after further development) could be evaluated
as part of the next Decadal Survey.  This 12 m flagship facility, the
``High Definition Space Telescope'' (\HDST) was described in our
AURA-sponsored report ``From Cosmic Birth to Living Earths''
(hereafter ``\CBLE''; Dalcanton et al. 2015).  \HDST\ would be a
general observatory, open to the entire astronomy community, that
would provide observational capabilities unmatched by any
concurrent facility from the far ultraviolet to the
near-infrared. \HDST\ would provide the combination of angular
resolution, stability, and photometric sensitivity with which a broad
community of researchers can make transformational discoveries in
planetary science, astrophysics, and cosmology. \HDST\ would take the
next leap in our ability to study energetic phenomena in the
ultraviolet (UV) with 50 to 100 times the sensitivity of the {\it Hubble
Space Telescope}. The UV is the regime with the most abundant and
most important astrophysical diagnostic spectral lines, and thus HDST would
expand our knowledge
of baryonic processes in the Universe tremendously,
particularly when combining its sensitivity with its spatial
resolution. No UV capability anywhere close to that provided by
\HDST\ is on the long-range plans for any major space agency. In
both UV and visible wavelengths, \HDST\ will probe main sequence
stellar populations into the centers of galaxies, map objects in the
outer Solar System with 20 to 200 km resolution, and provide
astrometric precision for many sources down to at least the microarcsecond
level. Such astrometric precision will open whole new
regimes of observational parameter space. We hope the community will
view \HDST\ as a facility for all of astronomy -- one that will open
many new windows into the cosmos.

In addition to its general astrophysics capabilities, \HDST\ would also be equipped to carry out the first large spectroscopic survey of habitable zone exoplanets.  By combining advanced starlight suppression with exceptional telescope stability, \HDST\ could survey hundreds of nearby stars for potentially habitable worlds, and then spectroscopically characterize the many dozens that are expected to host water on their rocky surfaces.  This capability would come on line in the mid-2030s, and would capitalize on -- and dramatically expand -- the important work done with the {\it Transiting Exoplanet Survey Satellite} ({\it TESS}), the {\it James Webb Space Telescope} (\JWST), the {\it Wide-Field InfraRed Survey Telescope} ({\it WFIRST}), ESA's {\it PLAnetary Transits and Oscillations} mission ({\it PLATO}), and 30 m class telescopes in the intervening decades.

Recently, Elvis (2015) posted a commentary on aspects of \HDST's exoplanet science case, focusing on \HDST's unique ability to search for biosignature gases around dozens of habitable planets. He addresses a number of important issues including whether the exoplanet characterization science could better be done from the ground, or with other space-based facilities, or with other target stars, or whether \HDST\ could meet its science goals at all (for either technical or budgetary reasons).  While all of these issues are addressed in \CBLE, we present a concise discussion of these issues below.  In particular, we take care to correct a number of misconceptions and errors that, if left uncorrected, could impair valuable community discussion of \HDST.

\section{Why Facilities for Direct Imaging \& Spectroscopy of Rocky Exoplanets Must Include a Large Space-Based Telescope}

Ground-based telescopes -- especially those expected to be operating
in the 2030s -- bring superb capabilities to the study of exoplanets.
The two main ground-based techniques that could potentially discover
biosignature gases, however, have limitations that make a space-based
direct detection mission compelling.  Transit spectroscopy is also not
a substitute for direct imaging and spectroscopy, although it is a
powerful complementary technique. We discuss these techniques in turn.

\subsection{Direct Detection \& Spectroscopy of ExoEarths from the Ground}

The large (20-40 m diameter) ground-based telescopes of the future aim to directly image planets orbiting in the Habitable Zones (HZs) of nearby bright M dwarf stars. Compared to other types of stars, planets around M dwarfs are thought to have the highest contrast ($\sim10^{-8}$ for M5's, and $10^{-7}$ for later types) at the NIR wavelengths where adaptive optics offer the best performance. Even though there are nearby M dwarfs, their habitable zones will be less than 0.04$''$ from the host star, requiring a $\sim$30 m telescope operating at its diffraction limit to observe the planets directly. However, given the expected performance of these telescopes, the number of M dwarfs that can be observed is likely to be small.  For current best estimates of likely coronagraphic performance, a dozen late M dwarf stars should be accessible\footnote{The likely number of target stars is still being debated in the community.}, and if higher contrast can be reached, dozens of earlier-type M stars may also be searched. Furthermore, the higher the desired contrast, the brighter the adaptive optics guide star must be to provide enough photons to sense the wavefront to the needed accuracy in less than an atmospheric coherence time.  This M star brightness limit (roughly K= 3 or V= 8) will further limit the number of targets accessible to ground-based telescopes. 

Ground-based coronagraphic observations will be an exciting first look
at the spectra of exoplanets in reflected light. Yet, only a fraction
of these may be rocky planets in the habitable zone, and only a
fraction of those may host life. If life and habitable zone planets
are common around M stars, then one of these dozen stars may indeed
yield a detection of a habitable world with biosignature gases. Such a
discovery would be profound, and would also indicate that life was
common, given the small number of stars surveyed. Both of these would
be strong immediate justification for a mission that explored the
nature of these biosignatures across a much larger population of
planets around a diversity of stellar types; M dwarfs have a number of
extreme properties that may make life on M dwarf planets unusual (as
we discuss in more detail below in Section~\ref{habsearch}), making a
broader survey compelling.  In the perhaps more likely event of a
non-detection of a biosignature gas, a larger sample would be
necessary, and could still yield a significant harvest of habitable
worlds, given that $\eta_{\rm life}$ would still have a 5\% (i.e.,
2$\sigma$) chance of being as large as $\sim$30\% if only a dozen
habitable worlds were surveyed.

In contrast, a 12 m \HDST\ would survey more than 500 stars, rather than dozens (\CBLE\ \S3.5.2 and \S3.5.3). This means that even a non-detection of a inhabited exoplanet with \HDST\ would still provide a significant constraint on our knowledge on $\eta_{\rm life}$, while also transforming our understanding of habitable zone planetary atmospheres. Furthermore, \HDST\ would probe complementary environments by surveying exoplanets around a range of stellar types (see Section~\ref{habsearch} below), rather than just cool M stars.

\subsection{Doppler-shift Spectroscopy for Atmospheric Characterization}

The Doppler shift technique of Snellen et al. (2010) cross-correlates
high-resolution spectra against template atmospheres.  As applied to
extremely large ground-based telescopes operating with 
starlight suppression (e.g., Snellen et al. 2013), this technique
requires many transits to build up to the needed signal-to-noise. It
can therefore only be used for extremely bright stars. While this is a
promising technique, it is not expected to be useable for more than 5
to 10 target stars.

There is also a possible technical sticking point for this
Doppler-shift technique. The wavelength-dependent speckles that remain
after starlight suppression would be 10,000 to 100,000 times brighter
than a planet that was $10^{-8}$ to $10^{-10}$ times fainter than its host
star. These speckles will not be easily removed, given that they may
not be uniform, stable, or measured well enough to calibrate them out
to the level needed.

\subsection{Is Transit Spectroscopy a Sufficient Replacement for Direct Imaging \& Spectroscopy?}

Transit spectroscopy with the {\it Spitzer Space Telescope} and the
{\it Hubble Space Telescope} (\HST) have provided the first spectra of
exoplanet atmospheres.  As exciting as these observations are, transit
spectroscopy probes the upper atmosphere of a planet, where the gas is
optically thin along the line sight. In contrast, direct spectroscopy
detects light that penetrates to much greater depths, before
reflecting off clouds and/or the surface of the planet. Compared to
transit spectroscopy, direct reflectance spectroscopy is sensitive to
features produced by the surface of the planet itself and has far more
access to the low altitudes where habitability indicators (such as
water vapor) and biosignature gases may be more abundant and more
likely to be found\footnote{For example, Earth's stratosphere is
  ``dry" because water is cold-trapped in the troposphere, and thus
  would not show water in a transit spectroscopic observation.}.

That said, it is not impossible that transit spectroscopy could detect
habitability signatures and biosignature gases that existed at high
enough altitudes.  In the near future, the best prospect is for
\JWST\ to use transit spectroscopy to study nearby exoplanets detected
in the next generation of transit surveys.  The upcoming NASA mission
{\it TESS} (Ricker et al. 2014; launch 2017) is expected to deliver 50
transiting rocky planets. A small number of those 50 will both be in
the star's HZ and be suitable for spectroscopic observations with
\JWST.  Current estimates suggest that roughly 5 or 6 habitable zone
transiting exoEarths are expected from {\it TESS} (Sullivan et
al. 2015). The follow-up atmospheric studies with \JWST\ are likely to
take several to tens of transits to build up sufficient
signal-to-noise ratios ({\it e.g.,} Deming et al. 2009). The vast
majority of these systems are expected to be around M dwarfs, which
have the highest probability of showing transits due to the closeness
of their habitable zones to the host star.  The ESA mission {\it
  PLATO} (launch 2024) will have a somewhat larger collecting area
than {\it TESS} and a more {\it Kepler}-like observing strategy,
giving it more sensitivity to habitable zone planets around hotter
stars. {\it PLATO}'s longer mission lifetime (6 years) will mean it is
better suited than {\it TESS} for robustly finding terrestrial sized
exoplanets in the habitable zones around F and G stars. Unfortunately,
transmission spectra of Earth-like atmospheres against sun-like stars
are not expected to be observable as not only is the signal too small
against the area of a sun-like star, but co-adding transits at yearly
intervals is impractical (Kaltenegger \& Traub 2009). In addition,
given {\it PLATO}'s launch date, it is not clear that \JWST\ will be
available to spectroscopically follow-up all of {\it PLATO}'s M star
exoplanet discoveries, although 30 m class telescopes potentially
could.

\subsection{Summary: A Direct Imaging and Spectroscopic Space Mission is Necessary}

For both of the above ground-based techniques, the search for biosignature gases could only be carried out for a handful of stars. These efforts could potentially lead to exciting individual discoveries, but only a large space telescope guarantees access to a more diverse sample of exoplanets that is tens to 100 times larger. Likewise, transit spectroscopy offers exciting opportunities to characterize the upper atmospheres of habitable zone exoEarths, but, again, the samples are likely to be small and almost entirely limited to M dwarfs. {\it WFIRST} will pursue direct detection experiments, but their sensitivity will be limited largely to brighter planets with better contrast from their host stars; spectroscopy of even super-Earths is likely to be out of reach, for all but one or two of the closest stars (Robinson et al. 2015). We also note that the accurate measurement of absorption features in the spectrum of an Earth-like exoplanet atmosphere will be challenging to do with ground-based telescopes, especially at low spectral resolution, given that the same features are present in Earth's own atmosphere and the widths of many of the Earth's spectral features are far broader than the expected Doppler shifts due to exoplanet orbital motions.

\section{Is the \HDST\ Exoplanet Observing Strategy Optimal?}

Elvis (2015) raises a number of concerns about the exoplanet observing strategy adopted in \CBLE. These issues include whether a focus on G stars is optimal, given likely progress using studies of M dwarfs with other facilities, whether the strategy of looking for biosignature gases in dozens of habitable zone planets was sufficiently sound to warrant a flagship mission, and whether the astrophysical unknowns are currently too large to consider a mission of this size.  We now discuss each of these issues in turn.

\subsection{What spectral types should a search for habitable planets include?}\label{habsearch}

Given that exoplanet spectroscopy is in its infancy, any spectrum of
an exoplanet atmosphere is valuable.  Even when restricted to
habitable planets, there is no stellar type for which a spectrum would
be uninteresting.

For ground-based and transit studies, essentially all habitable-zone
spectroscopic targets will be orbiting M dwarfs -- not because M
dwarfs' planets are intrinsically the most interesting cases to study
or the most likely to harbor life, but because they are the optimal
targets given the observational capabilities.  Likewise, for \HDST,
the distribution of spectral types observed is set entirely by
optimizing the design reference mission (DRM) described in Stark et
al. (2014, 2015).  The nominal \HDST\ DRM is agnostic about what
spectral types are observed\footnote{The only stars explicitly
  excluded from the \HDST\ DRM analysis are known close binaries
  ($<10$ arcsec separation), stars beyond 50 pc, stars with luminosity
  classes I, II, or III (i.e. giant or supergiant stars off the main
  sequence), or stars without an entry in the {\it Hipparcos} catalog
  (Perryman et al. 1997).  The latter is somewhat biased against K and
  M stars beyond 15 pc, but most K and M stars beyond this distance
  would be too faint for HZ exoplanet detection and
  characterization.}, and instead chooses an observing strategy that
maximizes the number of systems that can be detected and
spectroscopically characterized during a fixed amount of time.

\begin{figure}
\begin{center}
\includegraphics[width=0.95\textwidth]{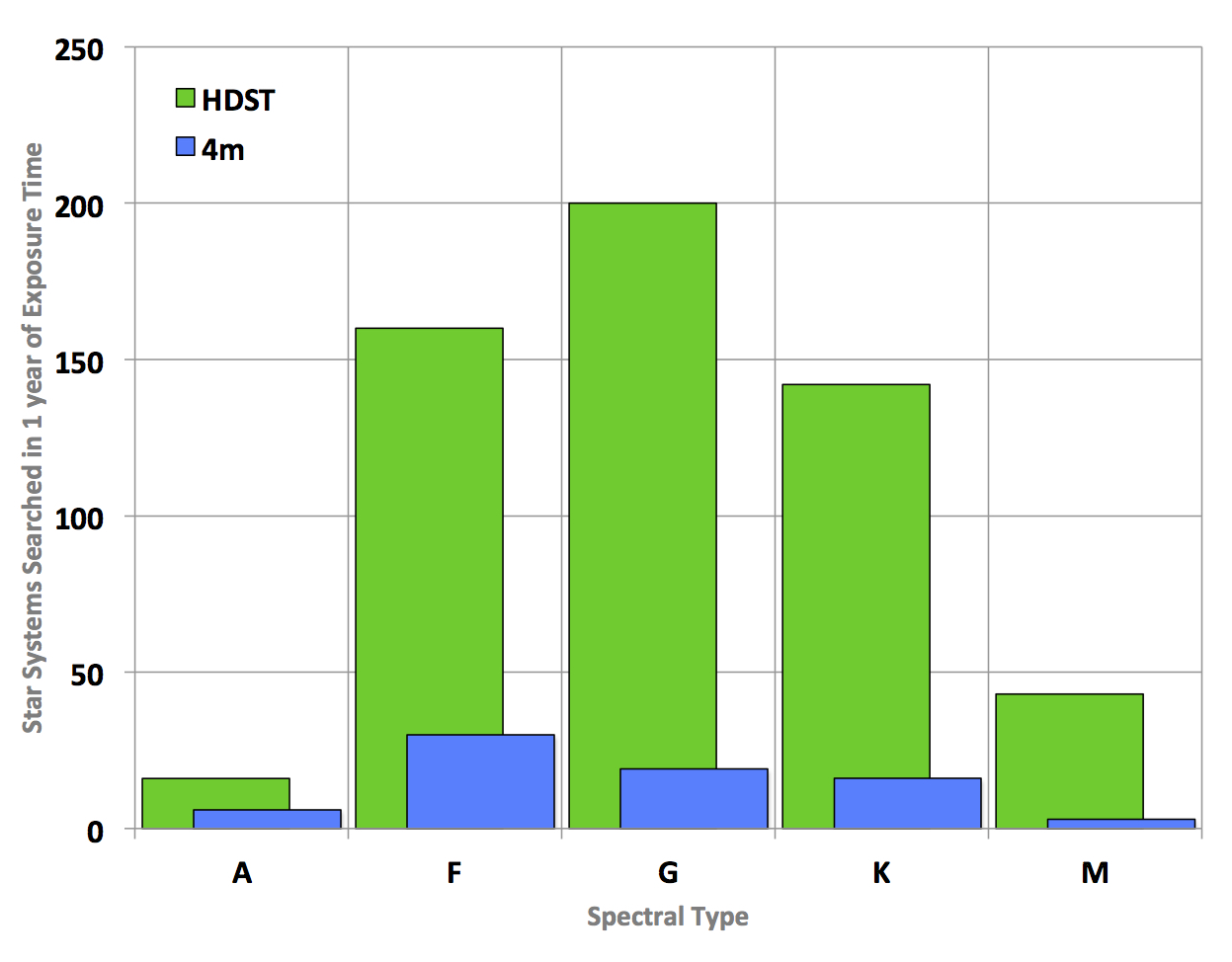}
\end{center}
\vskip-0.25cm
\caption{\small{The distribution of stellar spectral types in the design reference mission (DRM) for a 12 m \HDST\ (green) and for a 4 m exoplanet mission (blue). The histograms show the number of stars in each spectral type surveyed with a survey integration time of 1 year (which includes spectroscopic characterization of Earth-like planets). \HDST\ surveys a total of 561 stars. The 4 m space telescope surveys 74 stars. The Stark et al. (2015) altruistic yield optimization algorithm is used in these DRMs. }}\label{sptypes}
\end{figure}

For the specific case of 1 year of on-target integration time (approximately 2 years of mission time) with \HDST, the 561 stars that could be characterized span spectral types from A through M, as shown in Figure~\ref{sptypes}.  This distribution is in contrast with Elvis (2015)'s assertion that \HDST\ would observe only G-type stars.  The distribution does peak for G-type stars, but 64\% of the targets would be of spectral types other than G. Thus, while G stars offer the exciting opportunity to find a true Earth-twin, they dominate the \HDST\ sample solely because they are efficient for a space-based observatory to observe (\CBLE\ \S3.5).

It is likely that \HDST's broad coverage of spectral types may offer a better chance to find conclusive evidence of biosignature gases than a survey solely focused on M dwarfs.  M dwarfs' high rates of flare activity (West et al. 2015), coronal mass ejecta, and particle acceleration may make the environment in their habitable zones less conducive to the long-term viability of life on a planet's surface.  These same events can also complicate interpretation of any detected biosignature gases, because they may drive a planet's atmosphere out of equilibrium.

The community has a long list of other reasons to mitigate the risk of focusing solely on M stars. These include: high X-ray luminosity (Schmitt \& Liefke 2004) and EUV luminosity and high energy flares, all of which could be detrimental for life or more importantly could erode atmospheres ({\it e.g.,} Scalo et al. 2007); thin atmospheres on tidally-locked planets in HZ, which may be problematic for heat redistribution around the planet ({\it e.g.,} Joshi, Haberle and Reynolds 1997; Wordsworth 2015); and a higher luminosity phase of pre-main sequence evolution, which may promote water loss or even entire loss of atmospheres for rocky planets in HZ of M stars ({\it e.g.,} Luger and Barnes 2015).  In short, while Elvis (2015) correctly points out that the temperature differences between M and G stars are ``not so large," M stars are qualitatively different in many ways that have a direct impact on planetary atmospheres and habitability.  None of these are reasons not to look at M stars, but they are excellent reasons not to solely look at M stars.

While none of these concerns strictly precludes life ever evolving around an M star, they do suggest several reasons why the likelihood of life may be suppressed.  Given that only a few dozens of M stars can be searched from the ground or with transit spectroscopy (vs. $>$ 500 stars of many types with \HDST), and that not all of these will have planets in their habitable zones, it seems optimistic to assume that studies of M star planets are likely to find signs of life before the launch of \HDST. If, in the fortunate event that secure biosignatures were found around an M star in a search of dozens of stars, it would suggest that life was extremely common.  This discovery would make exploring habitable zones around stars spanning a broad range of spectral types even {\it more} compelling, not less, increasing the scientific value of \HDST.

\subsection{What are the uncertainties in the interpretation of biosignature gases?}\label{biosig}

Elvis (2015) questions \CBLE\ assertion that a large sample can help mitigate concerns that an individual detection of a biosignature gas may not actually have a biotic origin.

The astronomical search for life will employ a range of gases to
reduce the ambiguity of a single biosignature to acceptable
levels. For \HDST's wavelength range, these include H$_2$O, O$_2$,
O$_3$, CO$_2$, CH$_4$, and CO. The presence/absence of each of these
gases will help inform the case for a true biosignature
detection. Additionally, these detections will be placed in the
context of other information obtained on the planet's atmospheric
structure, the presence of clouds and aerosols, and the spectral
energy distribution of the host star. If the host star is flaring -
we'll know it. Much of this contextual information will likely be
absent from ground-based exoplanet atmosphere observations that cannot
observe in the optical (because they rely on NIR adaptive optics), or
from transiting exoplanet spectra that can only probe the very upper
atmosphere.  In particular, \HDST\ would be capable of characterizing
the UV signatures of energetic radiation that can modify the
atmospheric photochemistry. Such observations would then allow
possible false positive biosignature gases to be modeled
self-consistently.

Robust biosignature detections will rely on evidence that an atmosphere is out of equilibrium, or that one particular gas is extremely out of equilibrium with the rest of the atmosphere. Atmospheric disequilibrium can also be caused by short term stellar effects -- such as frequent high-energy flares ({\it e.g.,} Segura et al. 2010) or a runaway greenhouse ({\it e.g.,} Luger \& Barnes 2015) -- leading to false positives for biosignature gases. Many of these transient effects are most common around M dwarfs, making false positive biosignatures more likely in these late-type stars ({\it e.g.,} Tian et al. 2014; Harman et al. 2015; Luger \& Barnes 2015). Yet, if multiple systems, across a range of spectral types, show comparable signs of disequilibrium, then the odds go down that every single one of these systems is experiencing a transitory state.

\subsection{What is a scientifically useful sample size for habitable zone exoplanet spectroscopy?}

Elvis (2015) questions the sample size for ExoEarth yields in the
\CBLE\ report, properly quoting its carefully stated goal: ``to seek
dozens of exoEarths for detailed atmospheric characterization''. This
goal is to {\emph{seek}} biosignatures, not to find them, because it
is impossible to offer a guarantee that they will be found, even if
only at some arbitrary level of confidence.  This is why the concept
for the \HDST\ mission is not scoped to deliver biosignature
detections at a particular statistical significance.  Each
astronomer can also decide for themselves the definitions of ``common" or
``rare," and the point at which upper limits become uninteresting. Our
report contemplates these binomial statistics not to finely tune the
required aperture, but to show: (1) that a search of hundreds of stars and
an exoEarth sample of dozens has a reasonable chance of discovering
biosignatures even if they occur on only $\sim$10\% of habitable
worlds; (2) that \HDST's chance of doing so is much greater than smaller
missions fighting the $\sim$D$^2$ scaling of the exoEarth yield
(\CBLE\ \S3.5.3); and, not least, (3) that understanding why Earth-like
planets do not show biosignatures is almost as important as whether or
not they do. As shown in Chapter 3 of \CBLE, the \HDST\ sample of
dozens of exoEarths from hundreds of systems is the largest practical
search given our place in the Galaxy, the already known properties of
planetary systems, and the limits of foreseeable technology.

To understand the context of what might seem like small samples, and
the risks of investing counting statistics with too much weight, one
can consider another history-changing astronomical discovery. Edwin
Hubble's 1929 velocity-distance relation showed the expansion of the
Universe, but contained only 24 individual measurements of
Cepheid-based distances to nearby galaxies. Hubble's paper states
``For such scanty material, so poorly distributed, the results are
fairly definite''. Indeed. Yet Hubble's own estimate of his famous
constant was high by nearly a factor of 10 because of systematic
errors in the calibration of Cepheids that Hubble himself could not
have anticipated at the time. These errors were only identified and
resolved because Hubble's own modest sample showed the way, and proved
its importance. In hindsight, it would have been a mistake for Hubble
to have waited to embark on his program until he could obtain a
$3\sigma$ measurement, or dodge all systematics, or promise to deliver
an ``interesting'' null result.

Just as Hubble was the first astronomer to possess the tools to
measure the cosmic expansion, \HDST\ will be the first mission with
realistic prospects of imaging enough planetary systems to find and
characterize dozens of exoEarths. A single discovery of a biosignature
around an M star, which may come with the next generation of
ground-based measurements, would indeed be a great discovery. However,
\HDST\ would still be the only realistic prospect for expanding this
single detection by more than an order of magnitude, and placing this
first detection into context while reframing the big questions that
would motivate still larger samples in the future. Time will tell if
nature cooperates to deliver a detection among \HDST's dozens of
exoEarths, but we can be confident now that \HDST\ offers the most
exhaustive search of the nearby Galaxy that can be undertaken with
foreseeable technology -- a good practical definition of where to set
our scientific ambitions.

\subsection{Should we plan missions when there are astrophysical unknowns? Would we be better off waiting until after all observations of M dwarf planets are complete?}

One of the major conclusions of Elvis (2015) is that \HDST\ should not
move forward in 2035 because there will be progress from the study of
exoplanets in the habitable zones of M dwarfs in the meanwhile.  We
too look forward to this progress, even if limited to modest sample
sizes and a focus on M stars.

Given that the field will advance in the interval before \HDST\ launches, the important questions are whether the design of the \HDST\ mission is likely to fundamentally change as a result of what will be learned during the interval, and whether it will be as compelling in 2035 as it is today?  Alternatively, is the exoplanet community supportive of delaying the start of planning for a large exoEarth direct detection mission until the 2030s (for a launch in the mid 2040s, 30 years from now), to avoid the risk that an \HDST\ mission that started in the 2020s would not be the correct one?

\subsubsection{Would \HDST\ at its current size \& capabilities still be compelling in 2035?}

In the future that Elvis (2015) envisions, by 2035 we will know far
more about habitable zone exoplanets around M dwarfs than we do now,
with a reasonably good chance that one of the habitable zone planets'
atmospheres will show signs of biosignature gases.  Although we do not
consider the second of these assertions to be likely, we can still ask
if the design or size of \HDST\ would change, and if the arguments in
favor of \HDST\ would be weakened, if Elvis' (2015) future vision
indeed came to pass.

The first obvious thing that could potentially change is \HDST's
aperture. The current 12 m mirror diameter was chosen to allow
\HDST\ to survey hundreds of stars and take spectra of many dozens of
habitable zone exoplanets, assuming the Stark et al. (2015) survey
strategy.  The baseline Stark et al. (2015) model makes a number of
assumptions about astrophysical parameters (the probability $\eta_{\rm
  earth}$ that stars have a rocky planet in the habitable zone, the albedo
of the planet, and the exozodiacal background) and about starlight
suppression performance (primarily the achievable contrast and inner
working angle). The model then calculates a ``yield" of habitable zone
exoplanets that can be observed spectroscopically for a given
telescope diameter, as part of a large imaging survey of hundreds of
target stars.  When these parameters are held fixed, larger apertures
will always yield bigger samples, with the number of habitable zone
planet spectra scaling roughly as the square of the
diameter\footnote{Note that advances in starlight suppression and
  telescope design can also make \HDST\ yields more
  favorable. However, these are technical issues that will primarily
  be solved by investing in \HDST\ technology, and are largely
  independent of the {\it scientific} advances that may be made in the
  study of M dwarf habitable zone planets.  We therefore do not
  discuss these parameters further, and refer the reader to the
  already extensive discussions in \CBLE\ Chapters 5 and 6.}.

It may be that if the Universe can be shown to have more favorable
astrophysical parameters, then \HDST\ could reach the same yield with
a smaller aperture.  Some of the uncertainties in these parameters
will probably decrease in the coming decade.  Measurements of the
exozodiacal background are underway, although they will not
necessarily constrain its brightness at visible wavelengths in the
habitable zone itself. Likewise, {\it TESS}, {\it PLATO}, and
ground-based studies should improve measurements of $\eta_{\rm
  earth}$, but the uncertainties for F and G stars are likely to
remain high.  Measurements of planetary albedos of habitable zone
exoplanets will also be improved, but only for a handful of planets
around M dwarfs.

The reason these parameters will be uncertain at the launch of \HDST\ is that {\it only a large space mission is capable of measuring them reliably for stars warmer than M dwarfs}.  There is no suite of intervening experiments that could eliminate all these uncertainties, and thus there will always be a drive for larger apertures as the single most effective hedge against the Universe being unfavorably biased against easy detection of exoplanets.  

If there are no upcoming astrophysical measurements that seem capable of dramatically changing the scale of \HDST, are there scientific discoveries that would make the project scientifically less compelling?  In one possible future, the M dwarf searches for habitable zone planets fail to uncover any biosignatures, in which case the need for dramatically larger samples -- which only \HDST\ can deliver -- becomes all the more compelling. In the other, more dramatic future, the M dwarf studies do turn up evidence for biotically interesting atmospheres, in spite of the small sample sizes. In that case, the study of exoplanet life will be the new frontier. It defies reason to believe that the first hint of true ``astrobiology" will simultaneously end interest in the field. The history of astronomy demonstrates time and time again that the first discovery is never the last. Instead, it is the birth of a new field. 

\subsubsection{Would delaying \HDST\ until the mid-2040s be prudent?}

If, as Elvis et al. (2015) asserts, \HDST\ ``is not the right choice for NASA astrophysics at this time," then perhaps it could be the right mission, as long as it was delayed.  

We disagree that a delay is necessary, or that it would be advisable. The arguments for \HDST's basic size and capabilities are firmly grounded in the laws of physics -- the scaling of starlight suppression performance with telescope diameter and wavelength, the brightnesses of stars and their locations in the local neighborhood, and the sizes of habitable zones for stars of different luminosities and temperatures.  These same physical constraints will continue to hold in the future, and thus if the exoplanet community wishes to have more than few well-studied habitable zone exoplanets, a mission comparable to \HDST\ will be the only way forward\footnote{Barring rapid maturity of space interferometry coupled with large apertures, which seems unlikely at the present time.}.

Given that the design and launch of \HDST\ will likely take more than a decade, and that the laws of physics that drive its design will not change, there is no obvious scientific reason to delay developing such a mission.  Scientifically, \HDST\ will continue to be compelling, since it will remain the only path that leads to significant samples of spectra of habitable exoplanets.  Even with another decade of observations behind them, the 30 m class ground-based telescopes are unlikely to ever observe more than a few dozen M dwarfs.  Alternatively, an intermediate-sized space mission comparable to but somewhat larger than {\it WFIRST} -- with costs presumably also somewhat larger than {\it WFIRST}'s -- could serve as a half-measure that would launch in 2035 instead of the larger \HDST.  However, the fundamental scalings between yield and telescope diameter will limit such a mission to small sample sizes, comparable to the sample sizes of M dwarfs, but for warmer stars.  Such a mission would be several billion dollars, but with a yield of only a few spectra of habitable zone planets, it is inevitable that more questions would be raised than answered, and a mission like \HDST\ would need to follow.  If the concern is that exoplanet flagships risk Òcrowding outÓ a diverse suite of other flagships, this two-step path forward would lead to {\it two to three decades of exoplanet flagships being continuously under development}. We consider this less desirable than building \HDST\ -- which can potentially function for decades -- and complementing it with flagships at other wavelengths during its lifetime, reproducing the successes of the Great Observatories era.

Finally, rather than building an intermediate-aperture space mission, one could simply delay \HDST\ for another decade, and be content with the progress on M dwarf exoplanet systems from currently planned projects for the next thirty years.  While this may be the choice that comes out of the decadal survey process, it is hard to ignore the fact that exoplanet research is one of the most exciting, rapidly growing subfields of astronomy.  The number of refereed papers dealing with exoplanets has consistently more than doubled every 5 years, from 42 in 1999, to 202 in 2004,  to 474 in 2009, and 1185 in 2014, and these papers have been accessed via ADS over seven million times.  In the face of this explosive growth, strong community support for a flagship -- sooner rather than later -- seems likely.  The fact that \HDST\ {\it simultaneously serves the same large general astrophysics community as currently uses} \HST\ makes waiting thirty years unappealing as well.

\section{Are the technical goals of \HDST\ sensible?}

Elvis (2015) questions a number of technical aspects of the mission outlined in \CBLE. These include the actual needed observing time, the use of the optical for spectroscopic characterization, the achievable contrast, the choice of aperture, and the likely budgetary profile.  Here we correct a number of problems in the assumptions made by Elvis (2015).

\subsection{How much observing time is needed?}

Elvis (2015) does not properly describe the observing times needed for \HDST's program of exoplanet characterization. \HDST's program to spectroscopically characterize any habitable zone exoplanets found around $\sim$550 stars is based on the rigorous, refereed analysis presented in Stark et al. (2015).  This analysis uses the actual catalog of nearby stars and assumes a well-designed experiment that accounts for planetary phases, revisits required to thoroughly search accessible HZs, and the exposure time required to spectrally characterize each detected exoEarth candidate and search for O$_2$. When applied to \HDST, the Stark et al. (2015) framework suggests that in 1 year of on-sky integration and 2 years of mission time, \HDST\ will be able to image 561 nearby stars across many spectral types (Figure~\ref{sptypes}) to identify habitable zone planets, search the resulting $\sim 50$ planets for evidence of liquid water (assuming $\eta_{\rm earth} = 0.1$), and then thoroughly characterize the atmospheres of the subset found to be truly habitable.

In contrast, Elvis' (2015) Section 7 presents an approximate estimate of the observing time that would be needed for \HDST's exoplanet studies. This estimate incorrectly assumes that the \CBLE\ report ignores the time devoted to spectroscopy.  This estimate also does not adopt a reasonable plan for how such spectroscopy would be done. Elvis (2015) assumes that every star searched would have a HZ planet completely characterized at every wavelength, instead of the more sensible approach of gradually devoting increasing amounts of observing time to the most promising candidates. It also misuses \CBLE\ Figure 3-17, which is an exceptionally deep imaging example, to estimate a ``typical" exposure time, whereas the Stark et al. (2015) uses the {\it actual} observing time needed for each target star.  In short, the Elvis (2015) arguments in his Section 7 are not a correct analysis of observing time.

\subsection{Why focus on the optical and the near-IR, instead of the mid-IR?}

Elvis (2015) questions why \CBLE\ limits itself to exoplanet spectroscopy in the optical and near-IR (0.3 - 2.5 microns), correctly noting that there are also valuable atmospheric diagnostics at shorter and longer wavelengths. 

There are at least four straightforward, technical reasons to focus the first large survey of HZ exoplanets on the optical and near-IR: 
\begin{enumerate}
\item If one wishes to observe a habitable-zone planet at angle $\theta$ from its host star, reaching longer wavelengths requires a larger telescope ($\sim\lambda/$D). For nearby stars, moving out to the mid-IR would therefore require a currently inconceivably large telescope (or optical interferometer) that was more than 10 times larger than what would be needed at visible wavelengths.  
\item At longer wavelengths, only planets that have large angular separations from their host star can be detected. This limits samples to only the closest stars of any given spectral type. By the $\lambda/$D argument, visible wavelength observations significantly increase the sample of stars that can be observed, as compared to near-IR or mid-IR wavelengths.
\item The optical contains the wavelengths where HZ planets orbiting F, G, and K stars will be brightest in reflected light; exoEarths are expected to be an {\it order of magnitude} fainter in reflected light at 5 microns compared to 0.5 microns, while at the same time the surrounding zodiacal light is expected to be much brighter.  
\item Longer wavelength observations would require a cryogenically cooled telescope, which would add significant cost and complexity to an already challenging mission.
\end{enumerate}

As discussed above in Section~\ref{biosig}, there are many useful atmospheric molecules that produce absorption in optical and near-IR. \HDST\ should be able to detect spectral features of H$_2$O (a habitability indicator), O$_2$, O$_3$, CH$_4$, N$_2$O (traditional biosignature gases), CO$_2$, and CO, and other gases. Ongoing work aims to identify other potential biosignature molecules and evaluate spectra in the $<$ 2.5 micron range. The redder of these features will only be detectable in the closer stars, for which the habitable zones are (angularly) large enough to be studied at large values of $\lambda$ (Fig 3-16 of \CBLE). However, as noted in \CBLE, second generation coronagraphic instruments and/or starshades could allow longer wavelengths to be accessed for more stars, by improving the inner working angle of the starlight suppression system.

In summary, restricting an initial survey to long wavelengths would incur much greater costs for far fewer targets.  The research field of exoplanet atmospheric characterization (including exoEarths) will consume many generations of astronomers, just as the topics of stellar atmospheres and interiors, galaxies, quasars, and cosmology, have. Once terrestrial planets are discovered, later generations will aim to observe them at all wavelengths, as technology for formation flying, space interferometry, and on-orbit assembly come on line.

\subsection{What is the actual contrast needed for detection of an Earth-twin?}

Elvis (2015) misstates that the lower contrast ratio adopted by \CBLE\ for 1 micron spectroscopy and imaging would be inadequate to detect an Earth-like planet around a solar type star. To clarify, the $10^{-9}$ contrast level cited in \CBLE\ for NIR observations refers to the raw contrast value, which is the contrast achieved with the starlight suppression system during a science exposure. The final contrast value must indeed be $10^{-10}$, but this level can be achieved with post-observation processing.  A factor of 100 gain in the final contrast in post-processing has been demonstrated with \HST\ observations (Soummer et al. 2011; Rajan et al. 2015). This gain is accomplished by modeling the point spread function (PSF) by applying the Karhunen--Loeve image projection algorithm to a large library of PSF observations (Soummer et al. 2012). Factor of 10 or more improvements have also been achieved with ground-based direct imaging (e.g. Macintosh et al. 2015).
 
It is not unreasonable to expect that factor of ten gains will also be possible for the higher raw contrast ratios expected for future missions. Large ground-based telescope direct imaging programs currently envision reliance on post-processing gains.  In the near term, direct imaging post-processing techniques are currently being optimized for contrast levels of $10^{-7}$ to $10^{-8}$, as expected for the {\it WFIRST} coronagraph. Using simulated {\it WFIRST} coronagraphic data, and data from high-contrast testbeds (Ygouf et al. 2015) improvements of a factor of several are being obtained at the moment, and further progress seems likely. Hence, the \CBLE\ assumption that an additional factor of 10 gain in contrast can be achieved by improved PSF subtraction done subsequent to the observations is not unreasonable, and is fully consistent with current capabilities and assumptions used for the {\it WFIRST} coronagraphic mission. Whether or not post-processing or better raw contrast is required will continue to be a topic of ongoing research, but does not appear to be a technical show-stopper at this time.

\subsection{Is a 12 meter aperture needed?}

Elvis (2015) suggests that \HDST's aperture is unnecessary because its core exoplanet science mission can be done with other facilities on the ground, or that another (presumably much smaller) mission could reach equivalent science goals at 1/5th the imagined cost of \HDST, or that -- contradictorily --  even at 12-m, \HDST\ could not reach those goals either.        

We show elsewhere that the first and third of these assumptions are
not true.  As to the second, the exoEarth yield calculation in Stark
et al. (2015) makes very direct arguments as to why direct detection
and characterization of significant numbers of habitable planets
requires apertures comparable to \HDST's. The effect of aperture on
sample size can clearly be seen in Figure 1, where a 4 m UVOIR
telescope can only survey ~1/8th the number of planets as \HDST.
Larger apertures also offer more insurance against unfavorable values
of astrophysical parameters (like planetary albedo) that are simply
unknowable before launch. It is possible that some gains in technical
performance can potentially make somewhat smaller apertures more
competitive. Indeed, \CBLE\ argues for making the early investments
needed to accurately assess what performance can be
achieved for different telescope configurations operating in tandem
with possible coronagraph or starshade designs. Only after such
studies are done can the true, scientifically-acceptable lower limit
on aperture size be set.  We note, however, that there are no
technological advances that can make smaller apertures perform as well
for general astrophysics, as demonstrated extensively in Chapter 7 of
\CBLE.

\subsection{Can \HDST\ be built in the current budgetary climate?}

Elvis (2015) makes a number of comments about \HDST's cost and its relation to NASA's budget.  These points need clarification, as some of Elvis' (2015) claims are incorrect, and others are based on inaccurate assumptions.  Before tackling specific issues from Elvis (2015), we reiterate that \HDST\ is not a specific mission design whose cost can be accurately calculated without further development.  As discussed extensively in \CBLE\ (Chapter 7), taking \HDST\ from a mission concept to a true, budgeted mission will take further investment in technology and rigorous design studies.  Thus, all discussions of budget that follow should be considered as broad estimates and not firm numbers that can decide the fate of individual missions.

\subsubsection{What is the cost of \HDST?}

Elvis states that ``(\HDST's) large mirror surely drives the cost of
the mission''. However, \CBLE\ notes in Chapter 7 that
\begin{quote}
{\it While there is a rough D$^{1.6}$ parametric cost-scaling for the telescope itself, the scaling approach has only been established for small apertures and does not include the cost of the entire observatory, which would include instruments, testing, etc. For \JWST, the telescope itself was only $\sim$15\% of the total mission cost, leaving $\sim$85\% of the other observatory costs -- the spacecraft and sunshield, instruments, integration and cryogenic testing, ground system and operations, program management, and system engineering -- outside the realm where simple scaling models were valid.}
\end{quote}

\noindent Thus, while size certainly contributes to cost, one cannot
simply assume that \HDST\ must be more expensive than \JWST\ solely
because it is larger.

Elvis's subsequent argument -- that \HDST\ cannot fit in the existing
budget on a reasonable timescale -- is based upon cost assumptions for
\HDST\ that cannot be made without an actual, well-developed mission
concept.  While it certainly seems reasonable to assume that
\HDST\ would be significantly more expensive than {\it WFIRST}
(currently estimated at $\sim$\$2.5B, in real-year dollars), it cannot
be known if \HDST\ would be as high as \$10B.  As discussed in
\CBLE\ (\S7.2), \HDST's cost is likely to depend significantly on
factors that can in part be controlled by sufficient advance
technology investment and system engineering modeling.  Mission cost
depends strongly on the duration of the mission definition and
construction phases (Phases B, C and D), the level of tolerance built
into the system architecture, and the maturity of technology when the
mission goes into formulation. With sufficient initial investment (and
consistent with the next flagship mission wedge not opening up until
$\sim$2025), the cost for \HDST\ can be far better managed than was
possible with \JWST.

\subsubsection{Is there room in the NASA budget to build \HDST?}

Elvis misstates the total budget for NASA Astrophysics to be \$500M per year, whereas the FY2014 total NASA budget for astrophysics missions and related research is more than twice that (\$1.34 billion). This sum includes \$678M for the Astrophysics Division and \$658M for the James Webb Space Telescope project. This total is expected to be stable between FY2016 through FY2020, based on the FY2016 President's Budget Request. As the \JWST\ budget declines from its peak, which occurred in FY2014, that opening funding wedge is shifting almost completely back into NASA Astrophysics spending, where it is expected to remain for {\it WFIRST} and beyond. Materials presented\footnote{See http://sites.nationalacademies.org/cs/groups/ssbsite/documents/webpage/ssb\_168722.pdf} by NASA's Astrophysics Division Director, Paul Hertz, in August 2015 to the Space Studies Board suggest that,  after {\it WFIRST}'s launch in $\sim$2024, $\sim$\$580M per year would become available for future strategic astrophysics missions during the 2025-2035 decade. This leaves close to \$6B available for new flagships, assuming ongoing funding of $\sim$\$150M per year for Explorers and $\sim$\$80M per year for SOFIA during the same time period.  With participation from foreign space agencies, as much as \$7B could be available.  If enough technological maturation and system engineering can take place before 2025, building \HDST\ may not take the full 2025-2035 decade, which could reduce the overall mission cost.  In short, it is premature to accept Elvis' (2015) assertion that \HDST\ will ``take all of the available funding for 20 years."

\subsubsection{Can \HDST\ be built as part of a balanced program?}

Elvis (2015) states that \HDST\ would not permit any other astrophysics missions to be developed during its formulation and development phase. We agree that, without additional funding, it will not be possible for NASA astrophysics to simultaneously start two new flagship-class missions by 2025. But NASA will certainly be able to continue to operate whatever missions may be running at the time of an \HDST\ new start and would also be able to fund the vital Explorer program. Outside of the Astrophysics division, NASA will continue to build and operate a strong Planetary Science and Heliophysics mission portfolio as well. 

In addition, as a non-cryogenic telescope, there is nothing that prevents \HDST's operations for extending over decades, should it continue to be scientifically productive enough to justify the operations cost. It is therefore possible that \HDST\ will operate in tandem with future flagships -- from both the US and from other space agencies worldwide -- allowing us to return to the rich multi-wavelength capabilities we enjoyed during the era of NASA's ``Great Observatory" flagships.

\section{Summary}

The executive summary of Elvis (2015) lists seven strong conclusions critiquing specific aspects of \HDST's exoplanet mission. These are (verbatim):
 
 \begin{quote}{\it
(1) The focus on G-stars is not well justified; (2) only G-stars require the use of direct imaging; (3) in the chosen 0.5 -- 2.5 $\mu$m band, the available biosignatures are ambiguous and a larger sample does not help; (4) the expected number of biospheres is 1, with a 5\% chance of zero; (5) the accessible sample size is too small to give a 3 sigma upper limit that would show that exobiospheres are rare; (6) to get a sufficiently large sample would require a much larger telescope; (7) the extraordinarily rapid progress in the spectroscopy of planets around M stars - both now and with new techniques, instruments and telescopes already planned - means that a biosignature will likely be found in one before \HDST\ could complete its search in $\sim$2045. 
}\end{quote}

The analysis presented above raises significant questions about all of these top-level conclusions.  Taking these conclusions in turn, we find the following issues: 

\begin{enumerate}
\item The \HDST\ exoplanet science does not have an exclusive focus on
  G-stars, and there is clear scientific value in studying habitable
  zone exoplanets around more than just M stars.
\item Direct imaging is the most effective technique for exoplanet characterization, as it probes the entire atmosphere down to the planetary surface. No other techniques are as promising, especially for exoplanets with small ratios of atmospheric scale heights to planetary radii like those of Earth ($\sim$0.0013). Direct detection of exoplanets around stars other than M stars must be done from space.
\item \HDST\ will be sensitive to multiple atmospheric tracers that
  clarify the interpretation of biosignature gas features.  Larger
  samples reduce the chance that temporary atmospheric disequilibrium
  has led to a feature that mimics those of biotic origin.
\item The statistical significance of a null hypothesis is excellent when one has a sample size of at least 35 exoEarths. Reducing the uncertainty by a factor of nearly 20 on $\eta_{\rm water}$ or $\eta_{\rm life}$ (i.e., from $\pm$100\% down to $\pm$4\%) would be a major scientific advance. Smaller aperture missions cannot come close to achieving such strong constraints.
\item What is the definition of ``rare"?  Even for a 3$\sigma$ measurement, a sample size of 35 would reduce the uncertainty in $\eta_{\rm water}$ or $\eta_{\rm life}$ by a factor of $\sim$7, which still constitutes a dramatic improvement in our knowledge of the occurrence rate of habitable worlds\footnote{The sample size could only be marginally increased by extending the \HDST\ survey time because the exoplanet yield scales as just the 0.35 power of the survey time. Sample size could also be increased by installing a higher performance starshade or second generation coronagraph. Favorable astrophysical parameters could further increase the survey yield and the statistical significance of a null-detection.}, given that we know of exactly zero exoplanets with liquid water or life. Furthermore, focusing only on how well a single number -- $\eta_{\rm water}$ or $\eta_{\rm life}$ -- is known ignores the dramatic advances in scientific understanding that will come from a large direct detection survey of planetary atmospheres.
\item  Many reasonable definitions of ``a sufficiently large sample" are possible, each requiring a different minimum telescope aperture. Other groups could make well-justified arguments for both smaller or larger samples, given the lack of any constraints on either $\eta_{\rm water}$ or $\eta_{\rm life}$. \HDST\ is small enough that it can be built with current technology and launched with available rockets, but large enough that it would offer dramatic advances over {\it WFIRST} and 30 m class ground-based telescopes in the study of potentially habitable planets, while simultaneously revolutionizing astrophysics on all scales.
\item Prior to \HDST, progress in the study of habitable zone atmospheres will be limited by the small number of stars that can be studied. If these small samples of M stars are fortunate enough to reveal a single exoplanet with atmospheric biosignature gases before \HDST's launch, it will only serve to increase the interest in \HDST's study of dozens of such systems, across a range of host star types. To argue that the interest in \HDST\ would decrease after the detection of a biotically influenced atmosphere flies in the face of how science is done. Finding a prototype system has historically launched new fields of study, rather than finished them. Truly understanding habitable worlds beyond our own requires not just knowing their occurrence frequency, but also elucidating the physics and chemistry that controls their formation and evolution. To do that requires tens of systems to be studied.
\end{enumerate}

While much of this document has focused on clarifying particular points raised by Elvis (2015), it is vital to reiterate that these issues focus on a small fraction of the science that would be addressed by \HDST. While the search for habitable worlds and biosignature gases is a natural driver for a number of \HDST's technical choices, focusing solely on these misses the vastly larger scientific scope of the mission and its capabilities (UV sensitivity, integral field \& high resolution spectroscopy, wide field imaging, etc).

\HDST\ will do far more for the understanding of exoplanets than simply counting habitable worlds.  In addition, as the true successor to {\it Hubble}, \HDST\ extends {\it Hubble's} capabilities immensely in the spectral range where we can achieve extraordinary angular resolution.  \HDST\ can see the bulk of stellar populations, the strongest atomic lines, and, close to home, can monitor everything in the Solar System from Mars outwards. Virtually every astronomer in the world has used {\it Hubble} or its archives, or has capitalized intellectually from its scientific legacy. \HDST's impact and legacy is likely to be similar.

With that said, we also share Elvis' (2015) desire to maintain a broad space science program that maximizes the scientific return of NASA's investments. \HDST\ would bring important synergies with future observatories at many wavelengths and thus highlights the need for longer mission lifetimes (e.g., via on-orbit servicing) and well-funded national space science and human space exploration programs. We acknowledge that \HDST\ will be costly and may ultimately only be viable if ranked amongst the nation's highest priority research programs, as well as being highly ranked internationally. \HDST\ has the potential to be so ranked, but we firmly endorse that its scientific goals be evaluated as one of the many possible valuable astrophysics priorities for pursuit in the coming decade. 

\acknowledgements

The authors wish to thank a number of people for their input while crafting this response. These include Remi Soummer and Chris Stark from the Space Telescope Science Institute, Shawn Domagal-Goldman, Avi Mandell, and Karl Stapelfeldt from the Exoplanets and Stellar Astrophysics Laboratory at the Goddard Space Flight Center, Heidi Hammel from AURA, members of the Virtual Planet Laboratory, and colleagues at the University of Washington Department of Astronomy.

\end{document}